\documentclass[sigconf]{acmart}

\usepackage{todonotes}
\usepackage{enumitem}
\AtBeginDocument{%
  \providecommand\BibTeX{{%
    \normalfont B\kern-0.5em{\scshape i\kern-0.25em b}\kern-0.8em\TeX}}}
\setcopyright{acmcopyright}
\copyrightyear{2020}
\acmYear{2020}
\acmDOI{10.1145/1122445.1122456}

\usepackage{titlesec}
\newcommand\rurl[1]{%
 \href{http://#1}{\nolinkurl{#1}}%
}
\setcopyright{rightsretained}

\begin{document}
\copyrightyear{2020}
\acmYear{2020}
\acmConference[JCDL '20]{Proceedings of the ACM/IEEE Joint Conference on Digital Libraries in 2020}{August 1--5, 2020}{Virtual Event, China}
\acmBooktitle{Proceedings of the ACM/IEEE Joint Conference on Digital Libraries in 2020 (JCDL '20), August 1--5, 2020, Virtual Event, China}\acmDOI{10.1145/3383583.3398631}
\acmISBN{978-1-4503-7585-6/20/06}
\fancyhead{}
\title{Gandhipedia: A one-stop AI-enabled portal for browsing Gandhian literature, life-events and his social network}

\settopmatter{authorsperrow=3}

\author{Sayantan Adak*, Atharva Vyas*, Animesh Mukherjee}
\affiliation{%
  \institution{IIT, Kharagpur}
  \state{West Bengal -- 721302}
}

\if{0}\author{Atharva Vyas}
\email{atharvavyas@iitkgp.ac.in}
\affiliation{%
  \institution{IIT, Kharagpur}
  \state{West Bengal -- 721302}
}
\author{Heer Ambavi*}
\email{heer.ambavi@iitgn.ac.in}
\affiliation{%
  \institution{IIT, Gandhinagar}
  \city{Palaj}
  \state{Gujarat -- 382355}
}

\author{Animesh Mukherjee}
\email{animeshm@iitkgp.ac.in}
\affiliation{%
  \institution{IIT, Kharagpur}
  \state{West Bengal -- 721302}
  \postcode{}
}
\fi
\author{Heer Ambavi*, Pritam Kadasi, Mayank Singh}

\affiliation{%
  \institution{IIT, Gandhinagar}
  \city{Palaj}
  \state{Gujarat -- 382355}
}
\author{Shivam Patel}

\affiliation{%
  \institution{SRM Institute of Science and Technology}
  \city{Kattankulathur}
  \state{Chennai -- 603203}
}

\thanks{*These authors have contributed equally.}
\begin{abstract}
  We introduce an AI-enabled portal that presents an excellent visualization of Mahatma Gandhi’s life events by constructing temporal and spatial social networks from the Gandhian literature. Applying an ensemble of methods drawn from NLTK, Polyglot and Spacy we extract the key persons and places that find mentions in Gandhi's written works. We visualize these entities and connections between them based on co-mentions within the same time frame  as networks in an interactive web portal. The nodes in the network, when clicked, fire search queries about the entity and all the information about the entity presented in the corresponding book from which the network is constructed, are retrieved and presented back on the portal. Overall, this system can be used as a digital and user-friendly resource to study Gandhian literature. 
\end{abstract}

\ccsdesc[500]{Computing methodologies~Natural language processing}
\ccsdesc[500]{Computing methodologies~Network science}
\ccsdesc[300]{Computing methodologies~Ensemble methods}
\ccsdesc[500]{Information systems~Summarization}
\ccsdesc[300]{Information systems~Social networks}

\keywords{Entity recognition; Mahatma Gandhi; Temporal Networks; Text Processing}

\maketitle

\section{INTRODUCTION}
Mahatma Gandhi\footnote{https://en.wikipedia.org/wiki/Mahatma\_Gandhi} was the leader of India’s non-violent independence movement against British rule and in South Africa who advocated for the civil rights of Indians. Gandhi's life has been a major source of inspiration for many people. There is much written about his life, which are of great relevance to not only historians, but also general public interested in learning about his life and his ideologies. 

Compilation of these documents, and visual depiction of Gandhi's life could of unprecedented value to many people. \if{0}Significant efforts have been made in the past to make this literature easily available through digitization and open access to literature on various web portals.\fi

\noindent{\bf State-of-the-art}: Websites like \rurl{gandhiheritageportal.org}, 
\rurl{mkgandhi.org}, etc., contain major works of Gandhi, be it his letters, magazines, newspaper articles, or speeches in a digital format providing easy access to the literature on Gandhi. In addition, there are a good number of scholarly digital libraries, which have sizable collection of resources on Gandhi in digital forms\footnote{JSTOR, HathiTrust Digital Library, World EBook Library, South Asian Archives, Internet Archives, Project Gutenberg, Shodhganga, WorldCat.}. The abundance of literature on these platforms results in information overload and makes it difficult to gather relevant information as per a user's interest.  This is when life has become faster than ever and users tend to get disengaged very quickly.


\noindent\textbf{Contributions}: In this demo paper, we discuss the development of Gandhipedia, an interactive web portal that aims to digitize all the writings of Mahatma Gandhi and present them in a well organised format which can serve as a useful and easy-to-access resource for users. We have successfully constructed the spatial and temporal networks for the 7 key texts authored by Mahatma Gandhi. The nodes in these networks are entities (people or places that find mention in these texts) and edges correspond to co-mentions of entities within a pre-defined time window. The nodes in the network on click fire a query about the corresponding entity and search the text to return excerpts from different chapters in the text where the entity occurred. \if{0}The primary idea behind this visualisation is to keep users engaged and to feed relevant information to them incrementally as per user requirements. This is best explained by the following scenario.\fi Consider a user interested in all occurrences of ``Gopal Krishna Gokhale''\footnote{https://en.wikipedia.org/wiki/Gopal\_Krishna\_Gokhale} in the autobiography (The Story of My Experiments with Truth\footnote{https://en.wikipedia.org/wiki/The\_Story\_of\_My\_Experiments\_with\_Truth}) of Gandhi. The user can look into the network, quickly find the node ``Gokhale'' and click on it to obtain all information organised chapter/time wise about ``Gokhale'' without having to skim through the entire autobiography to find and manually aggregate this information. In addition, the portal also enables text based query search across 100 volumes of the Collected Works of Mahatma Gandhi. The current portal is hosted at \rurl{http://gandhipedia150.in}.

\label{lit}


\section{GANDHIPEDIA ARCHITECTURE}

\begin{figure}[!h]
\centering\includegraphics[scale = 0.22]{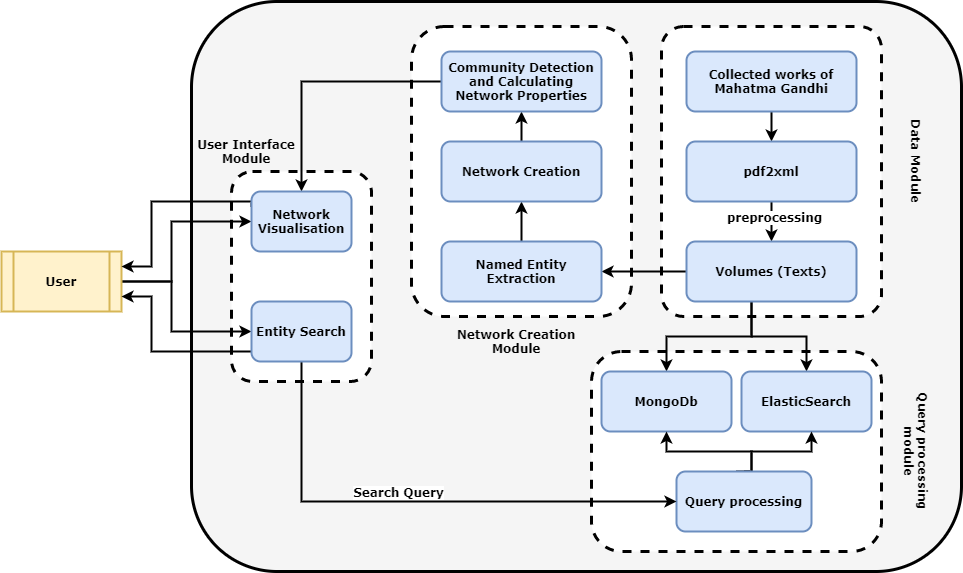}
\caption{Architecture of Gandhipedia.}
\label{gandhipedia_architecture}
\end{figure}

We leverage the Collected Works of Mahatma Gandhi (CWMG) available at \cite{CWMG}, a collection of 100 volumes of letters, speeches and books written by Gandhi for the construction of the portal. Figure~\ref{gandhipedia_architecture} shows the detailed Gandhipedia portal architecture.  It consists of four distinct modules: (i) data module, (ii) query processing module, (iii) network creation module, and (iv) user interface module. The arrows represent the direction of data flow. Next, we describe each module in detail.

\noindent \textbf{Data module}: We use pdf2xml tool\footnote{https://sourceforge.net/projects/pdf2xml/} to convert the volumes from PDF format to XML format. The XML files are processed to detect chapter boundaries. Each chapter refers to a book, letter, speech, or newspaper article. Each chapter is stored individually in text format in MongoDb\footnote{MongoDb: https://www.apress.com/gp/book/9781430230519}.

\begin{figure}[h]
\centering\includegraphics[scale = 0.29]{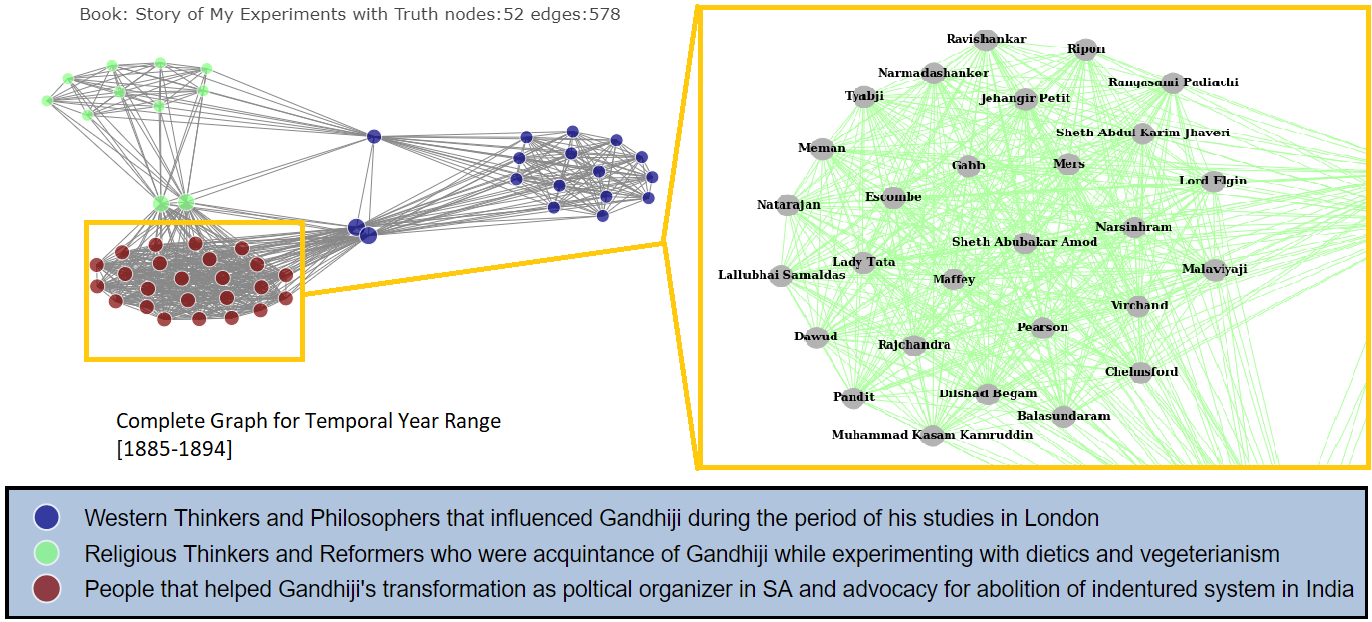}
\caption {Sample temporal people graph shown in web portal. The broad level interpretations of each of the communities are noted in a line of text below the network.}
\label{whole_newtork}
\end{figure}

\noindent \textbf{Network creation module}: We currently implement this module for seven books authored by Mahatma Gandhi, including his autobiography. An example temporal network is shown in Figure~\ref{whole_newtork}. Different colors represent different communities. The network construction methodology is described as follows -- (i) \textit{Identifying place/location entities}: We recognize named entities (place/person) occurring in different chapters of books. We use an ensemble of three different NER libraries, NLTK\footnote{https://www.nltk.org/}, Polyglot\footnote{https://draquet.github.io/PolyGlot/}, and Spacy\footnote{https://spacy.io/}. We only consider entities that were identified by at least two libraries. (ii) \textit{Filtering noisy entities}: Above NER mechanism results in several common nouns that were neither people or person. We filter out these entities using a list of cities and countries from NLTK. We use WordNet to filter improper nouns\footnote{https://wordnet.princeton.edu/}. (iii) \textit{Temporal networks creation and clustering}: Temporal network creation consists of three important steps. First, each chapter is mapped to a specific year. All entities in a particular chapter are associated with that year. All entities in a particular year are linked to each other. In addition, entities of year $t$ are linked with entities of year $t-1$  and year $t+1$. The hypothesis behind construction of such networks is that entities mentioned close by in time could possibly be `socially' related. Finally, community detection is performed using standard Louvain and Infomap clustering algorithms~\cite{louvain-infomap}. 
 
\begin{figure}[!tbh]
\centering\includegraphics[scale = 0.38]{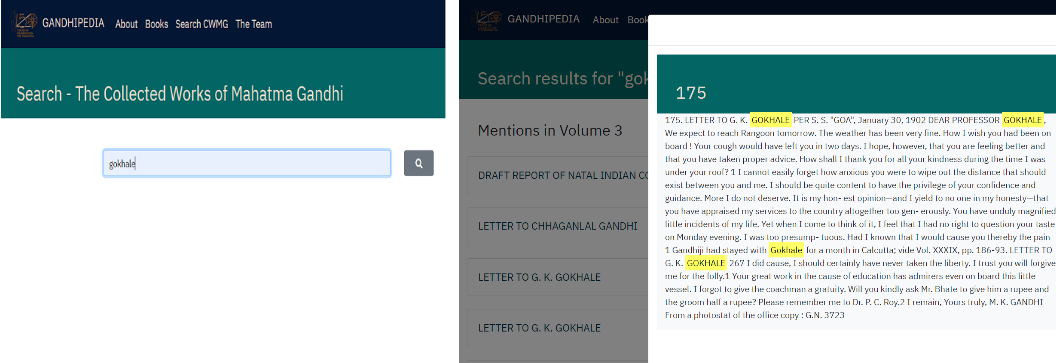}
\caption{\label{search_query}Sample search result shown on the web portal.}
\end{figure}

\noindent \textbf{Query processing module}:
The query processing module processes two distinct types of queries, network visualization related and full-text based search using ElasticSearch~\cite{10.1145/2597073.2597091}. \textit{Search through the network}: For the network visualization, when a node is clicked (person/place) a search query with the entity as the query term is fired. The search results are organised time wise and chapter wise and returned back to the user. \textit{Search through the CWMG}: The current full-text search supports first 40 volumes of CWMG in a textual format highlighting the query and its results as shown in Figure~\ref{search_query}. It supports general query retrieval using MongoDb database.

\noindent \textbf{User interface module}:
This module fetches results from the query processing and the network creation modules. The entity search displays the volumes, books and then chapters, with the searched entity highlighted. The network visualization\footnote{Dash: https://github.com/plotly/dash on top of Plotly: https://plot.ly.} depicts the generated interactive networks arranged over time. 

\if{0}
\section{SYSTEM DESCRIPTION}

We use Flask~\cite{10.5555/2621997} to handle server requests. The network visualization is achieved using Dash~\cite{Dash} on top of Plotly~\cite{plotly}. Gandhipedia supports general query retrieval using MongoDb~\cite{mongoDB} database. The full-text search facility is supported using  ElasticSearch~\cite{10.1145/2597073.2597091}.\fi

\section{CONCLUSION AND FUTURE WORK} 
We develop Gandhipedia to study Gandhian literature and his social networks in an interactive manner. In future, we plan to index more than 100 volumes of CWMG and present multilingual search facility. We also aim to develop timelines to represent which personalities/places are mentioned for the first-time in the literature. This helps in development of nice chronological visualisations.  

\if{0}\section{ACKNOWLEDGEMENT}
We would like to acknowledge the National Council of Science Museums (NCSM), the Gandhi Smriti and Darshan Samiti (GSDS) and the Ministry of Culture (MoC) for the continuous support throughout the project. We express our special gratitude to Shobhana Radhakrishna and Ravi Chopra for the very productive discussions.\fi

\bibliographystyle{ACM-Reference-Format}
\bibliography{gandhipedia}
\end{document}